\documentclass{jfm}

\usepackage{graphicx}
\usepackage{natbib}

\ifCUPmtlplainloaded \else
  \checkfont{eurm10}
  \iffontfound
    \IfFileExists{upmath.sty}
      {\typeout{^^JFound AMS Euler Roman fonts on the system,
                   using the 'upmath' package.^^J}%
       \usepackage{upmath}}
      {\typeout{^^JFound AMS Euler Roman fonts on the system, but you
                   dont seem to have the}%
       \typeout{'upmath' package installed. JFM.cls can take advantage
                 of these fonts,^^Jif you use 'upmath' package.^^J}%
      }
  \else
  \fi
\fi

\ifCUPmtlplainloaded \else
  \checkfont{msam10}
  \iffontfound
    \IfFileExists{amssymb.sty}
      {\typeout{^^JFound AMS Symbol fonts on the system, using the
                'amssymb' package.^^J}%
       \usepackage{amssymb}%
       \let\le=\leqslant  
         
      }{}
  \fi
\fi

\ifCUPmtlplainloaded \else
  \IfFileExists{amsbsy.sty}
    {\typeout{^^JFound the 'amsbsy' package on the system, using it.^^J}%
     \usepackage{amsbsy}}
    {}
\fi

\newsavebox{\astrutbox}
\sbox{\astrutbox}{\rule[-5pt]{0pt}{20pt}}

\newcommand{\khat}{\ensuremath{\mathbf{\hat{k}}}}
\newcommand{\uvec}{\ensuremath{\mathbf{u}}}
\newcommand{\adv}{\ensuremath{\uvec\cdot\nabla}}
\newcommand{\nuss}{\ensuremath{\mathrm{Nu}}}
\newcommand{\pran}{\ensuremath{\mathrm{Pr}}}

\title[Dynamics of fingering convection II]{Dynamics of fingering convection II: The formation of thermohaline staircases.}

\author[S. Stellmach, A. Traxler,  P. Garaud, N. Brummell and T. Radko]%
{S. \ns S\ls T\ls E\ls L\ls L\ls M\ls A\ls C\ls H$^{1,2,3}$,%
A.\ns T\ls R\ls A\ls X\ls L\ls E\ls R$^2$, \break
P.\ns G\ls A\ls R\ls A\ls U\ls D$^2$, \ns
N. \ns B\ls R\ls U\ls M\ls M\ls E\ls L\ls L$^2$,\break
\and T.\ns R\ls A\ls D\ls K\ls O$^4$}

\affiliation{$^1$ Institut f\"ur Geophysik, Westf\"alische
Wilhelms-Universit\"at M\"unster, D-48149 M\"unster, Germany \\[\affilskip]
$^2$Applied Mathematics and Statistics, Baskin School of Engineering,
University of California, Santa Cruz, CA 96064, USA\\[\affilskip]
$^3$ Institute of Geophysics and Planetary Physics,
University of California, Santa Cruz,CA 96064, USA\\[\affilskip]
$^4$Department of Oceanography, Naval Postgraduate School, Monterey, CA 93943, USA}
\pubyear{210}
\volume{???}
\pagerange{???--???}
\date{?? and in revised form ??}

\begin{document}

\maketitle

\begin{abstract}
  Regions of the ocean's thermocline unstable to salt fingering
  are often observed to host thermohaline staircases,
  stacks of deep well-mixed convective layers separated by thin
  stably-stratified interfaces. Decades after their discovery,
  however, their origin remains controversial. In this paper we use
  3D direct numerical simulations to shed light on the
  problem. We study the evolution of an analogous double-diffusive system,
  starting from an initial statistically homogeneous fingering state
  and find that it spontaneously transforms into a layered state. By analysing
  our results in the light of the mean-field theory developed in Paper I,
  a clear picture of the sequence of events resulting in the staircase
  formation emerges. A collective instability of
  homogeneous fingering convection first excites a
  field of gravity waves, with a well-defined vertical
  wavelength. However, the waves saturate early through regular but
  localized breaking events, and are not directly responsible for the
  formation of the staircase. Meanwhile, slower-growing, horizontally
  invariant but vertically quasi-periodic $\gamma$-modes are also
  excited and grow according to the $\gamma-$instability mechanism.
  Our results suggest that the nonlinear interaction between these
  various mean-field modes of instability leads to the
  selection of one particular $\gamma-$mode as the staircase
  progenitor. Upon reaching a critical amplitude, this progenitor
  overturns into a fully-formed staircase. We conclude by
  extending the results of our simulations to real oceanic parameter
  values, and find that the progenitor $\gamma-$mode is expected to
  grow on a timescale of a few hours, and leads to the formation of a
  thermohaline staircase in about one day with an initial spacing of
  the order of one to two metres.
\end{abstract}


\section{Introduction}
\label{sec:intro}

Fingering convection is a small-scale mixing process driven by a well-known
double-diffusive instability of stably stratified fluids.
When density depends on two components with opposite
contributions to the overall stratification, a
fingering instability may develop if
the diffusivity of the stably stratified component is larger than the
diffusivity of the unstably stratified one.
This commonly occurs in a large variety of natural environments, the best studied one being the
heat-salt system in the oceanic thermocline \citep{schmitt1994ddo,schmitt2005edm}.
Other areas of application range from the Earth's atmosphere \citep{merceret1977pmd,bois2000dda} to magmatic melts \citep{tait1989ccv} and the interiors of
giant planets \citep{guillot1999igp} and stars \citep{vauclair2004mfa,charbonnel2007tmp,stancliffe2007cem}.

Scientific interest in the dynamics of fingering convection, the saturated
state of the fingering instability, is twofold. When viewed as a small-scale
turbulent process, it is an intrinsic source
of diapycnal mixing in stably stratified regions, which operates even in the absence
of mechanical forcing. As such, it can play an important
role in controlling the global transport of heat and chemical species
in the system considered (see reviews by Schmitt, 1995 and \citet{ruddickgargett2003}).
Fingering convection is also known to drive dynamics on much larger scales by
exciting a variety of secondary instabilities, such as internal gravity waves
through the collective instability
\citep{stern1969cis,stern2001sfu}, thermohaline intrusions \citep{stern1967,toolegeorgi1981,walsh1995int}
and the more recently discovered $\gamma$-instability \citep{radko2003mlf}. Generally
speaking, these secondary instabilities are driven by a positive feedback mechanism
between large-scale temperature and salinity perturbations, and the turbulent
fingering fluxes induced by these perturbations. In Paper I,
we provided definitive
measurements of the small-scale turbulent fluxes induced by
fingering convection in the heat-salt system, and unified all previous
theories of the large-scale secondary instabilities in a single ``mean-field'' framework.
In this paper, we now turn to one of the longest standing unsolved problems related
to fingering convection, namely the formation of thermohaline staircases.

Thermohaline staircases are formed as stacks of deep convective
layers, both thermally and compositionally well-mixed, separated by
thin fingering interfaces.
They are ubiquitously associated with active fingering convection,
and have been observed both in laboratory experiments
\citep{stern1969sfa,krishnamurti2003ddt,Krishnamurti2009hsa} and in oceanic
field measurements \citep{tait1971ts,schmitt2005edm}. Oceanic staircases are
long-lived, and can span the entire thermocline, with individual
layer heights ranging from 10m-100m and horizontal extents of km-size or more.
Regions of the upper ocean which exhibit a staircase
stratification have much larger diapycnal mixing rates compared with smoother
regions supporting the same overall temperature and salinity contrasts
\citep{schmitt2005edm,veronis2007}.
The layering phenomenon therefore needs
to be understood to quantify its
role in the overall transport balance in the ocean, and
by extension, in many other systems as well.

Forty years after their original discovery, however, the mechanism
responsible for the formation of thermohaline staircases essentially
remains an enigma. Existing theories might
be roughly classified into two groups. In a first class of models,
layer formation is triggered by external effects in addition to
fingering convection. It has been speculated for example
that staircases and homogenous fingering both
represent distinct metastable equilibria of the same system,
and that layers are formed by external events which
cause vigorous mechanical mixing and drive the system from one equilibrium to the
other \citep{stern1969sfa}. Since staircases
are often observed in regions with significant lateral gradients
of heat and salinity (such as the Mediterranean outflow for example),
it has also been proposed that these gradients are actually required
to trigger their formation, through the excitation and nonlinear evolution
of the thermohaline intrusions \citep{walsh1995int,merryfield2000ots}.

By contrast, a second class of ideas favour the notion that layering
is inherent to the dynamics of fingering convection,
with no need for external forcing. In these models, staircase formation
is thought to arise from the nonlinear development of the
two other large-scale modes of instability mentioned earlier,
namely the collective instability \citep{stern1969cis,stern2001sfu} or the
$\gamma$-instability \citep{radko2003mlf}.
\citet{stern2001sfu} studied the linear and nonlinear evolution of gravity waves
excited by the collective instability, arguing that the overstable
waves would eventually overturn and break into layers.
However, while their numerical simulations show evidence for localized
breaking events, these do not result in layer formation.
\citet{radko2003mlf} on the other hand put forward the $\gamma$-instability as the
direct precursor of thermohaline staircases, since by nature
the normal modes of this instability are vertically modulated but
horizontally invariant perturbations in temperature and
salinity. His two-dimensional direct numerical simulations (DNS),
which are the first to exhibit unseeded spontaneous staircase
formation from an initially homogeneous fingering field, appear to
confirm his hypothesis.

Despite recent progress in this type of models, a number of fundamental questions remain
to be addressed. The first relates to the validity of the mean-field
framework: do the mean-field equations correctly model the large-scale
dynamics of fingering systems? This question has so far only partially
been answered for the linearised mean-field equations
through idealised tests in which a single large-scale
perturbation is seeded on the fingering field, its growth rate monitored
and successfully compared with theoretical predictions \citep{stern2001sfu,radko2003mlf}.
However, whether such perturbations are indeed
excited by the turbulence with a sufficiently large amplitude
to rise above the noise level, and with sufficient scale separation
for mean-field theory to be applicable, remains to be determined
in the general case.
Furthermore, we note that these tests have so far only
been performed in two-dimensional
numerical simulations. Given the well-known pathological nature of the
energy transport from small to large scales in 2D turbulence
\citep{kraichnan1967}, convincing validation of mean-field theory
can only come from 3D simulations.

The second question is related to the problem of layer formation itself.
Existing theories \citep{stern2001sfu,radko2003mlf} have each
put forward a particular mean-field mode of instability
as the precursor for staircase formation. As such, they consider
each mechanism in isolation and ignore other possible unstable modes.
In Paper I we identified the fastest growing modes, and showed that
in the parameter regime where staircases are observed
several kinds of mean field instabilities can simultaneously be excited. In fact modes on
a considerable range of spatial scales, both direct and oscillatory in nature, are predicted to grow by
mean field theory, and it seems likely that their mutual interaction disrupts their evolution.
The resulting dynamics remains to be explored.

Finally, all previously proposed models assume that the nonlinear evolution
of a mean-field mode eventually triggers the formation of layers.
Unfortunately, the predictive
power of these mean-field models breaks down as soon as turbulence on a
wide and continuous
range of spatial scales is generated. Three-dimensional computer simulations
resolving all relevant scales currently are therefore the only
way to study the final stages of layer formation.

The purpose of the present paper is to answer some of these questions, and
ultimately illuminate the problem of staircase formation, via
large-scale, three-dimensional numerical simulations of
fingering convection. Since both small-scale fingers as well
as deep layers have to be resolved simultaneously for this to be possible,
and since fingers and staircases evolve on vastly disparate timescales,
such simulations had until recently been impossible. Thanks to the
rapid advances in high-performance computing, this is no longer the case.

Our paper is organised as follows. We begin by briefly
summarising in Section \ref{sec:eqs} the basic equations
for fingering convection, and describe our numerical model setup
and selection of governing parameters.
In Section \ref{sec:sims} we present our results, the first
large-scale three-dimensional simulation
showing spontaneous layer formation, which is fully resolved down to the dissipative scale.
We analyse the simulation in the light of the mean-field theory presented in Paper I,
focusing on the evolution of the various large-scale modes of linear
instability. Our results enable us to place constraints on the
applicability of mean-field theory, and
lead to a clear and simple picture of the sequence of events finally resulting in
layer formation.
We conclude in Section \ref{sec:ccl} by summarising our results and by discussing
their implications both in the oceanographic context, and further afield.

\section{Equations and model setup}
\label{sec:eqs}

\subsection{Governing equations}
\label{sec:goveqs}

In natural systems, both fingering convection and staircase formation usually occur
far from physical boundaries. It is therefore important to select a
numerical setup which minimises boundary-effects.
Following \citet{stern2001sfu} and \citet{radko2003mlf}, for example,
we consider a fingering-unstable
system, permanently forced by background vertical temperature and salinity gradients
$T_{0z}$ and $S_{0z}$. In this case, the non-dimensional model equations are:
\begin{subequations}
\begin{eqnarray}
\frac{1}{\mathrm{Pr}} \left(\frac{\partial\uvec}{\partial t} + \adv \uvec\right) & = & -\nabla p + (T - S)\khat + \nabla^2 \uvec \mbox{   ,} \label{eq:momentum} \\
\nabla \cdot \uvec &=& 0 \mbox{   ,}\label{eq:continuity}  \\
\frac{\partial T}{\partial t} + w + \adv T & = & \nabla^2 T \mbox{   ,} \label{eq:heat} \\
\frac{\partial S}{\partial t} + \frac{1}{R_0}w + \adv S & = & \tau \nabla^2 S \mbox{   ,} \label{eq:composition}
\end{eqnarray}
\end{subequations}
where $T$ denotes perturbations away from the linearly
stratified background temperature, $S$ the similarly defined salinity perturbation,
$p$ the pressure perturbation from hydrostatic balance,
and the vector $\uvec = (u,v,w)$ is the velocity vector. The above
system has been non-dimensionalised using the
anticipated finger scale $d=(\kappa_T \nu / g \alpha T_{0z})^{1/4}$ \citep{stern1960sfa}
as the lengthscale (where $\kappa_T$ is the
thermal diffusivity, $\nu$ the kinematic viscosity, $g$ is gravity and $\alpha$ denotes the thermal expansion coefficient), and the thermal
diffusion timescale across $d$, namely $d^2/\kappa_T$, as the timescale.
Temperature and salinity have been rescaled  by $T_{0z} d$ and $\alpha
T_{0z} d / \beta$ respectively, where $\beta$ is the saline contraction coefficient. Three non-dimensional parameters control the behaviour of the system:
the diffusivity ratio $\tau =\kappa_S/\kappa_T$, where $\kappa_S$ is the salt diffusivity,
the Prandtl number Pr $= \nu/\kappa_T$, and the background density ratio
$R_0 = \alpha T_{0z} / \beta S_{0z}$.

To minimise the influence of boundaries, perturbations are assumed to be
periodic in all three spatial dimensions. Hence for a computational domain of
size $(L_x,L_y,L_z)$ we set $q(x,y,z,t) = q(x+L_x,y,z,t) = q(x,y+L_y,z,t) = q(x,y,z+L_z,t)$
for $q \in \{T, S, \uvec\}$. This configuration guarantees in particular
that layers cannot be
generated artificially by flux convergence at solid boundaries
\citep{ozgokmen1998nsl,merryfield2000ots}. It is also important to
note that the total horizontal averages of the temperature and salinity fields
are not assumed nor forced to be
zero, so that the mean vertical profiles of temperature and salinity can
evolve freely with time.

Finally, note that by contrast with Paper I,
we do not include lateral gradients, and thus suppress
the mechanism driving intrusive modes. As such
we restrict our analysis to spontaneously emerging staircases, and show
how they may form even in the absence of any additional forcing. Although
intrusions are important in regions of the ocean subject to very strong lateral gradients,  this assumption is reasonable in more typical regions of the thermocline since Paper I shows that for small-to-moderate
lateral gradients, intrusive modes grow on
much slower timescales than either gravity-wave modes or
$\gamma-$modes.

\subsection{Selection of the governing parameters}
\label{sec:govpars}

Simulations which have to resolve both fingers and possible
larger scale structures such as thermohaline staircases
still pose significant numerical challenges.
As seen in Paper I (see Table 1), for parameter values appropriate for
salty water ($\pran = 7,\,\tau \sim 0.01$), even a ``small domain
simulation'' containing only a few fingers requires a resolution of the order of
$1000^3$ for low values of the density ratio to be fully resolved. Since our goal
is to simulate a much larger domain, we are constrained to use a
larger value of the diffusivity ratio instead, and choose $\tau = 1/3$.
The Prandtl number on the other hand can be taken to be that of
water ($\pran = 7$) without difficulty. Guided by Radko (2003),
we adopt $R_0 = 1.1$ as the background density ratio.

\begin{figure}
\centerline{\includegraphics[width=\linewidth]{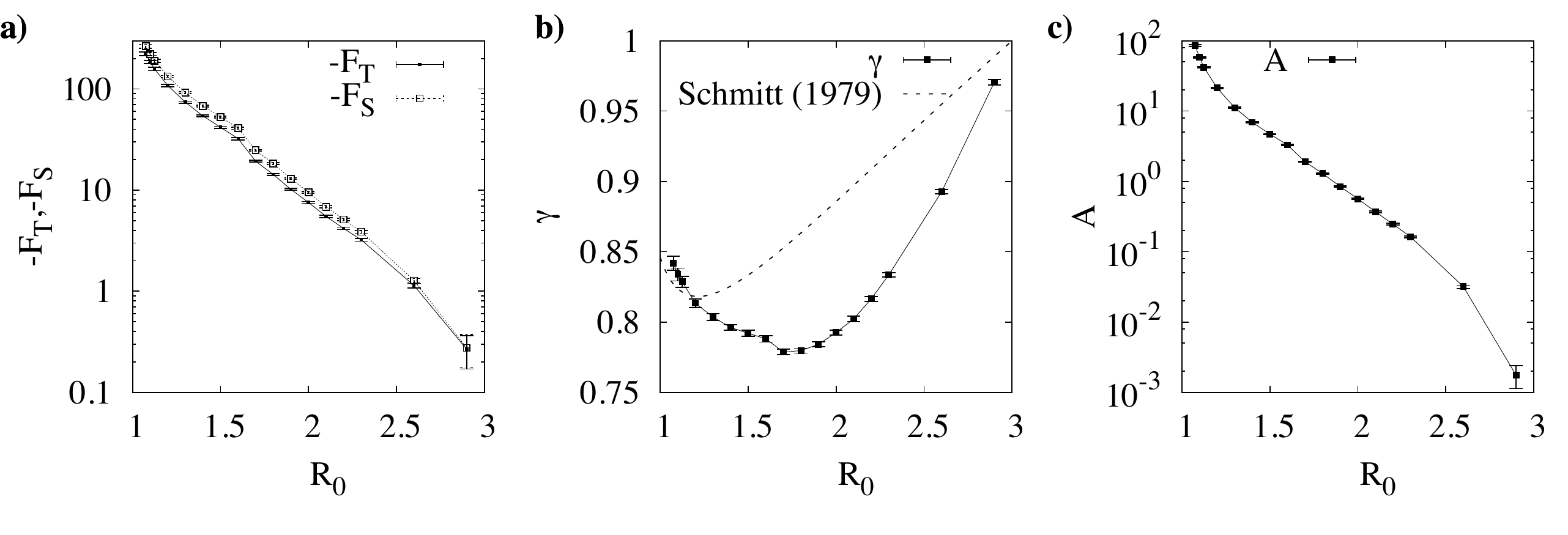}}
\caption{Numerically determined buoyancy fluxes due to heat and salt,
  $F_T =\langle wT\rangle$ and $F_S = \langle wS\rangle$ respectively, their ratio
$\gamma$, and the Stern number $A= (F_S - F_T )\,/\,\pran \,(1 - 1/R_0)$, as a function of $R_0$ for $\tau=1/3$ and $\pran =7$. The theoretical prediction of \citet{schmitt1979fgm} is shown in {\bf b)} for reference. The fluxes were measured
after integrating the model described in Section \ref{sec:goveqs} to saturation, in a small domain of size $5 \times 5 \times 15$ fastest-growing finger widths (FGWs, see \citet{schmitt1979fgm}). The experimental protocol for these measurements is similar to that described in the Appendix of Paper I. A necessary condition for the $\gamma$-instability \citep{radko2003mlf} is that $\gamma =F_T / F_S$ should be a decreasing function of $R_0$. As shown in Figure \ref{Fluxes_laws_Le_3}b, this happens here when $R_0 < 1.7$.
A necessary condition for the  collective instability \citep{stern2001sfu} is that the Stern number $A$ should be larger than an order one factor. As seen in Figure \ref{Fluxes_laws_Le_3}c this condition is satisfied only for $R_0 < 1.7$ as well.  }
\label{Fluxes_laws_Le_3}
\end{figure}

\begin{figure}
\centerline{\includegraphics[width=0.6\linewidth]{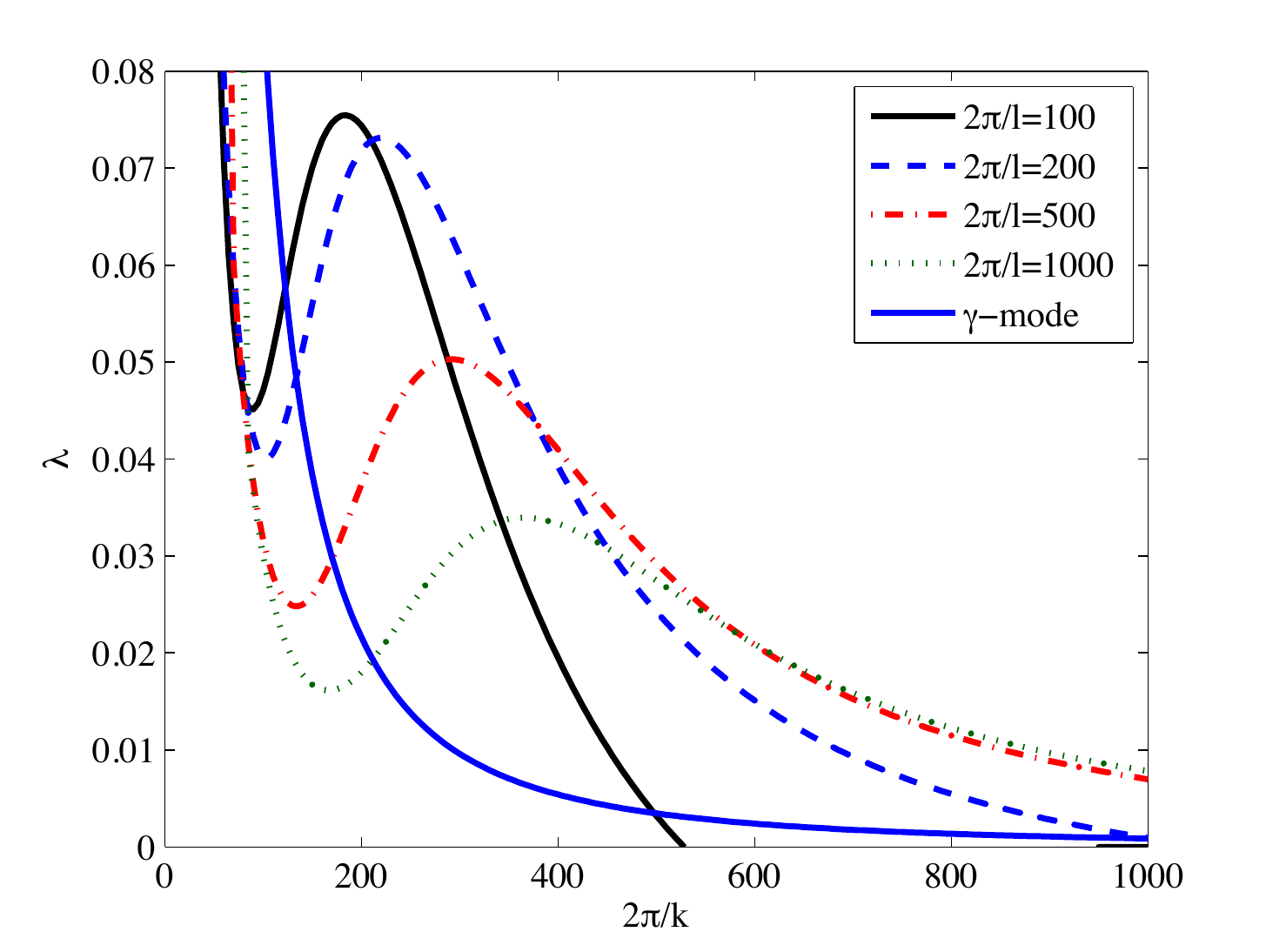}}
\caption{Growth rate $\lambda$ of mean-field instabilities for $\pran = 7$,
  $\tau=1/3$, and $R_0=1.1$, using the mean-field theory presented in Paper I and the
  flux laws presented in Figure \ref{Fluxes_laws_Le_3}. The horizontal axis represents the vertical wavelength in units of $d$, and the different curves represent different horizontal wavelengths as noted in the caption. Note that mean-field theory does not apply to structures which only contain a few individual fingers. This limits its validity to modes with horizontal and/or vertical wavelengths larger than about 100$d$ (see Section \ref{sec:phase2} for detail). }

\label{Theoretical_Growth_Rates_Le_3}
\end{figure}

As described in the introduction, we are interested in using the
numerical simulation to study the causal relationship between
the collective- and $\gamma$-instabilities and layer
formation. Oceanic staircases are observed in regions with low density ratio,
typically $R_0 < 1.8$ (e.g. Schmitt 1981). In this parameter regime,
mean-field theory predicts that gravity-wave modes grow much more rapidly than
$\gamma$-modes (see Figure 5 Paper I).  We must then verify that our selected numerical parameter values $\pran = 7$, $\tau = 1/3$ and $R_0=1.1$ yield the same
relative ordering of the growth rates of the various mean-field modes of instability.
As shown in Paper I, these growth rates are solutions of a cubic
equation where the coefficients of the cubic depend
on the governing parameters ($R_0$, $\pran$, $\tau$), on the spatial structure
of the mode, and on the functional dependence of the turbulent heat and salt fluxes on
the density ratio. Using the methodology outlined in Paper I, we
first measure these flux laws from small-domain experiments (Figure \ref{Fluxes_laws_Le_3}), and then
use the results to calculate the growth rates of the mean-field modes (see Figure \ref{Theoretical_Growth_Rates_Le_3}).

Figure \ref{Theoretical_Growth_Rates_Le_3} confirms that gravity-wave
modes and $\gamma$-modes are both unstable in our numerical parameter regime, providing us with the opportunity to study them simultaneously. Furthermore gravity
waves are expected to dominate the system's behaviour while the
$\gamma$-modes grow much more slowly.  This choice of parameters thus
guarantees that the   basic properties of the unstable modes are
comparable to those of the heat-salt system. Hence, despite the
difference in parameters, we anticipate that our
simulation should provide a reasonable view of the
dynamics of real thermohaline staircase formation at $\pran = 7$,
$\tau = 1/100$ and $R_0 \sim 1.5$.

\subsection{Numerical considerations}

Our large-scale simulation is performed
in a domain of size $335 d \times 335  d \times 536 d$, which corresponds to $27 \times 27\times 43$FGWs. The numerical algorithm for the solution of
(\ref{eq:momentum}-\ref{eq:composition}) with periodic
boundary conditions is based on the
standard Patterson-Orzag spectral
algorithm and is presented in more detail in Paper I.
Using small-domain tests to determine an adequate resolution for
this parameter regime, we settled
on a resolution of $576^3$ equivalent grid points for the large-domain run.
The computation was initialised with small amplitude white noise perturbations in
the temperature and salinity fields, and evolved from $t=0$ to $t=900$
(in dimensionless time units). Note that the initial resolution was sufficient for most of the simulation
but became inadequate in the layered regime, leading to a considerable accumulation of
energy in the high vertical wave-number modes of the salinity
field. Doubling the vertical resolution before layers start to form
solved this problem: the results were interpolated to a $576 \times 576
\times 1152$ grid slightly {\em before} the layering transition
and the simulation  was then re-computed on the finer grid from there on.

\begin{figure}
\noindent\centerline{\includegraphics[width=0.95\linewidth]{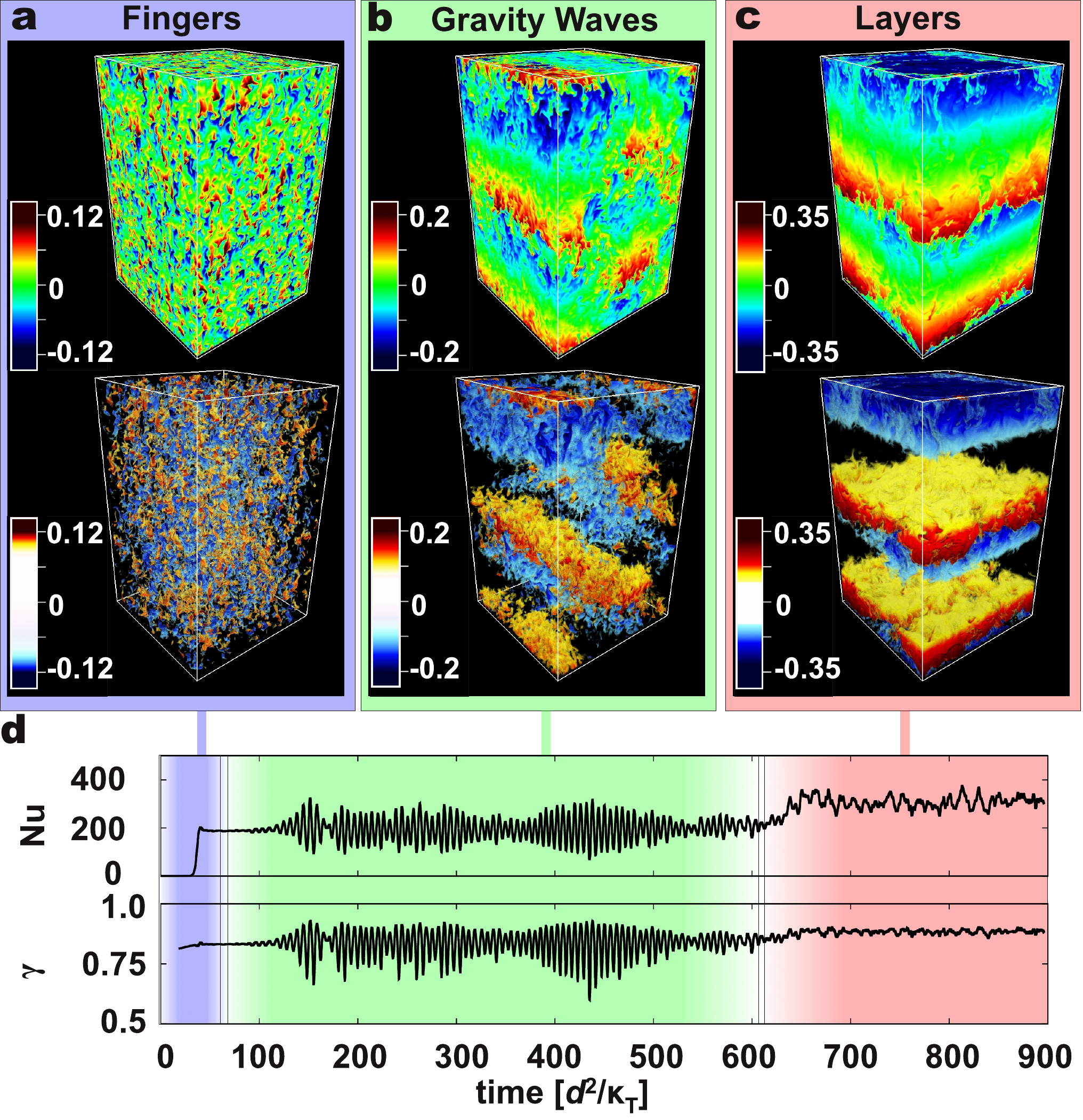}}
\caption{(a-c) Simulated layer formation in a large triply-periodic domain ($335 d \times 335  d \times 536 d$). Temperature perturbations normalised to the
  temperature difference across the whole domain are shown in
  colour, with their respective colour scales in each panel. The top
  panels are visualisations on the data-cube faces, while the lower
  ones are volume-rendered images.  Snapshots are shown
  for three characteristic dynamical phases. Fingering convection develops first (Phase I)
  and soon becomes unstable to large-scale gravity waves saturating at
  finite  amplitude (Phase II). Eventually, vigorously convecting
  layers form, separated by thin fingering interfaces (Phase III). (d) Time-series
  of the Nusselt number Nu$=1-F_T$ and of the turbulent flux ratio
  $\gamma = F_T / F_S$. }
 \label{figure_simulated_layering}
\end{figure}

\section{Direct simulation of staircase formation}
\label{sec:sims}

The dynamics of the system, as observed in our numerical simulation,
are  divided into three distinct phases illustrated in
Figure \ref{figure_simulated_layering}. Note that a movie of this simulation is
available in the online supplementary material. In Phase I,
initial perturbations are amplified by the
fingering instability, grow and eventually saturate into a state of
vigorous fingering convection ($0 \le t \le 100$;
Figure \ref{figure_simulated_layering}a). The global
heat and salt fluxes, measured by $\nuss=1-F_T$ and $\gamma = F_T /
F_S$, are found to be identical to those
determined previously in a smaller computational domain (see
Figure \ref{Fluxes_laws_Le_3}a). However, the system does not remain
in this homogeneous fingering state for long.
Gravity waves rapidly emerge, grow and saturate
(Phase II, for $100 \le t \le 600$), later followed by a sharp
transition to a layered state (Phase III, $t > 600$).
Phases II and III are now analysed in more detail.

\subsection{Phase II : gravity-wave phase}
\label{sec:phase2}

Shortly after saturation of the fingering instability ($t \sim 100$),
large-scale oscillatory modulations of the fingering field
appear. They rapidly grow until $t = 150$ when they, in turn, also
saturate. We argue below that these modulations are large-scale
finite amplitude gravity waves excited by the collective instability.

The visualisation of the temperature field presented
in Figure \ref{figure_simulated_layering}b and in the online movie,
for $t > 150$, reveals the
presence of a dominant perturbation pattern, with a horizontal
wavelength commensurate with the box width ($335 d$),
and a vertical wavelength half the box height ($268d$).
Its temperature amplitude at saturation is of the order of 15\% of the background temperature
difference across the domain, and equivalently 30\% of the temperature difference
across one wavelength. The spatial pattern and relative amplitudes of the density
and salinity perturbations are similar.

Figure \ref{figure_simulated_layering}d reveals
that these large-scale perturbations strongly affect
the global heat and salt fluxes. Both $\nuss$ and $\gamma$
begin to show visible oscillations around their respective background
values (i.e. as achieved in the homogeneous fingering
state) shortly after $t=100$, with increasingly larger amplitudes
until saturation at about $t=150$. By analogy with the effect of a pure
gravity wave on global heat transport, we expect the Nusselt number
to oscillate at twice the wave frequency, and similarly for $\gamma$.
A power spectrum of the Nusselt number
reveals a strong peak at $\omega \simeq 1.04$, so we estimate the wave
frequency to be $\omega_g \simeq 0.52$.
In our dimensionless units, the theoretical oscillation frequency for a pure
gravity wave with phase planes inclined at an angle $\theta$ from the
vertical is $\omega_g = \sqrt{\pran (1-1/R_0)} \cos{\theta}$, in other words
$\omega_g =  0.8 \cos{\theta}$ for our selected parameter values.
The dominant frequency observed is therefore compatible with the presence of a gravity wave with
$\theta=49 ^\circ$, which is consistent with our qualitative observation of the wave geometry.
These results also confirm the
theoretical prediction that gravity waves are the most rapidly growing
mean-field mode in this parameter regime, and suggest that their
absence from the 2D simulations of Radko (2003) is an artefact of reduced
dimensionality\footnote{The difference in the wave dynamics in 2D and 3D may be attributed to the stronger negative feedback of the wave shear on salt fingers in 2D.  In three dimensions, salt fingers tend to align into salt sheets parallel to the shear flow, which only moderately affects their vertical transport.  In 2D such alignment is geometrically impossible and therefore adverse effects of shear on salt-fingers, and thus on the collective instability, are more profound.}.

\begin{figure}
\centerline{\includegraphics[width=\linewidth]{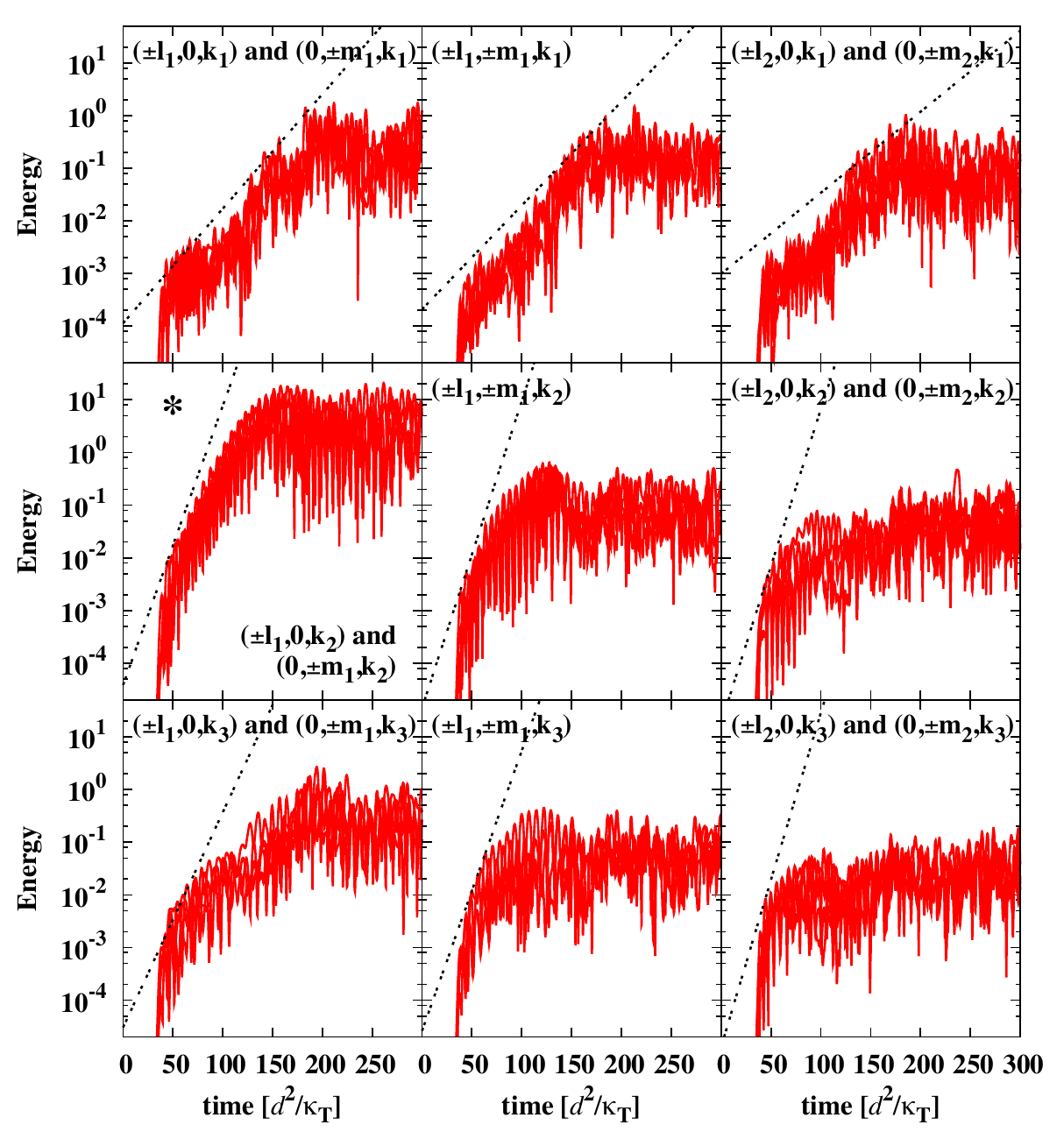}}
\caption{Kinetic energy in the first nine quartets of gravity wave
  modes, compared with the growth rate predicted using the mean-field
  theory presented in Paper I. Owing to the horizontal symmetry of the
  problem, modes $(\pm l,\pm m,k)$ and $(\pm m,\pm l,k)$ all have the
  same predicted growth rate, and are superimposed in each panel. The
  vertical wavenumber increases from top to bottom and the horizontal
  wavenumber increases from left to right. The $(\pm l_1,0,k_2)$ and $(0,\pm m_1,k_2)$ modes (denoted by a star) are the ``dominant modes'' referred to in the main text.}
\label{compare_gwave}
\end{figure}

We now study the gravity waves more quantitatively in the light of the
linear mean-field stability analysis presented in Paper I.
In this triply-periodic system, all mean-field modes have the normal
form $\hat q_{lmk} \exp(ilx+imy+ikz)$ where $\hat q_{lmk}$ is the mode amplitude, so that their predicted evolution can be compared
straightforwardly with Fourier-transforms of the output of the simulation.
Using this method, we extract from the numerical results
the amplitude of each velocity component $u$, $v$ and $w$, and compute
the kinetic energy $E_{lmk} = (|\hat u_{lmk}|^2+|\hat v_{lmk}|^2+|\hat w_{lmk}|^2)/2$ of any mode with wave-vector ${\bf k} = (l,m,k)$, as a function of time. The results are
shown in Figure \ref{compare_gwave}.

For simplicity, in what follows, we use the notation $l_n= 2n\pi /L_x$,  $m_n= 2 n \pi/L_y$,
as well as $k_n = 2n\pi/L_z$. Hence the dominant gravity-wave modes
identified by eye in Figure \ref{figure_simulated_layering}b, which have one wavelength in the
horizontal direction and two in the vertical direction, are denoted by ${\bf k} = (l_1,0,k_2)$ or ${\bf k} = (0,m_1,k_2)$.
Figure \ref{compare_gwave} shows that this set of modes, identified by a star symbol, begins to grow exponentially as soon as the background
fingering convection reaches a saturated state. By $t = 200$, their
kinetic energy is ten times larger than any of the
other modes.  Comparison of the observed growth rate with mean-field theory
shows reasonable agreement from $t=50$ to $t=150$, although is
slower than predicted by about 30\%.

\begin{figure}
\centerline{\includegraphics[width=0.95\linewidth]{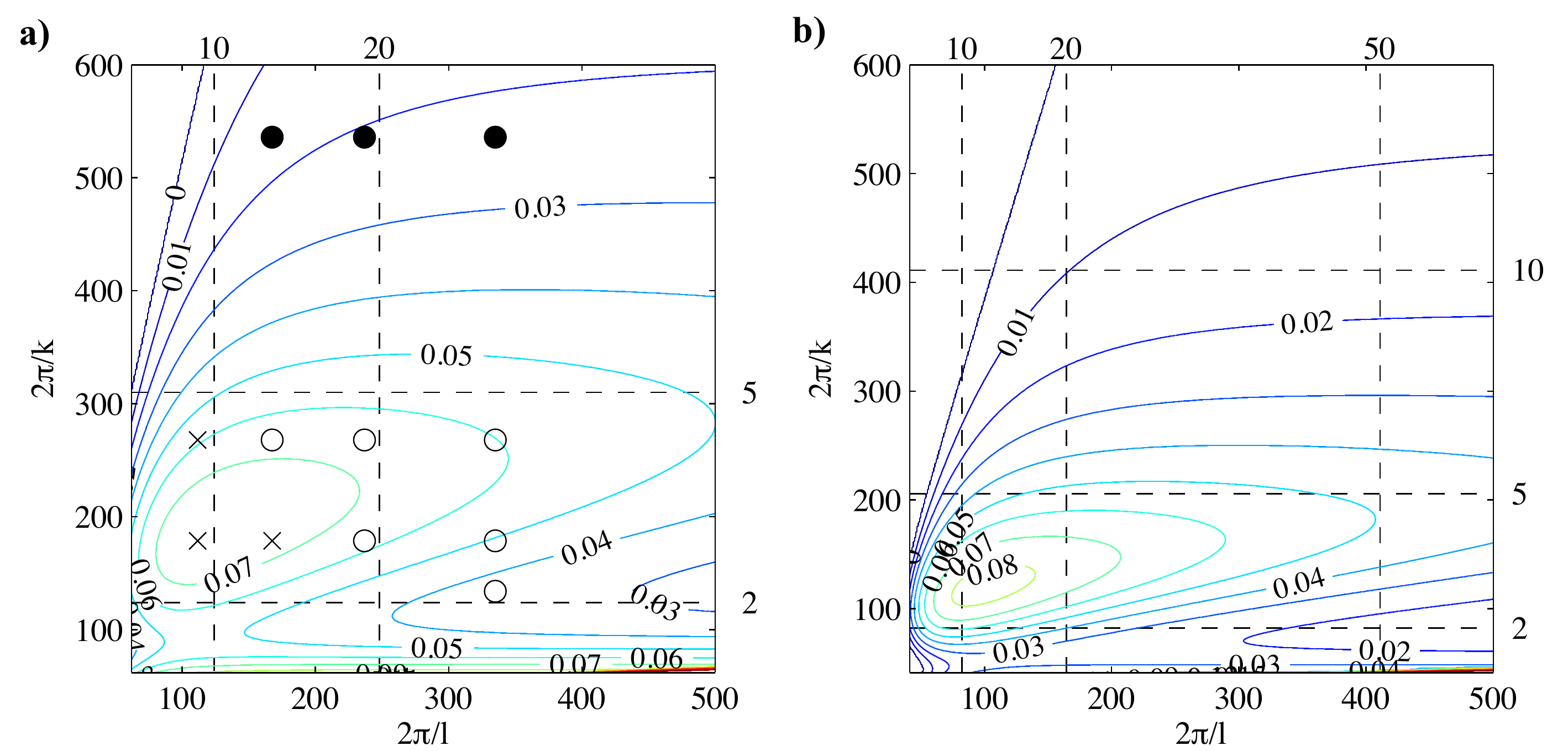}}
\caption{Contour plot of the growth rate of mean-field modes for a)
 our selected parameters $ \pran = 7$, $\tau=1/3, R_0=1.1$ and b) the heat/salt case $\pran =7, \tau=1/100$ with $R_0=1.5$. The individual modes presented in Figure \ref{compare_gwave} are identified on the left plot by symbols,
showing modes which grow with the expected growth rates as filled dots, modes which grow but more slowly than expected as empty dots, and modes which do not grow at all as crosses. The individual modes, from top to bottom, have $k = k_1$, $k=k_2$, $k=k_3$, and $k=k_4$, and from right to left, have $(l=l_1,m=0)$, $(l=l_1, m=m_1)$, $(l=l_2,m=0)$, $(l=l_3,m=0)$ and $(l=l_4,m=0)$.
The horizontal and vertical lines show the size of the structures
considered in terms of finger widths and heights (taking, as shown in Paper I, that
fingers are about 1 FGW wide and 5 FGW tall).
The equivalent plot for the heat/salt system shown on the right, which was obtained by using
the turbulent flux laws presented in Paper I, is very similar when viewed in terms of finger sizes.
}
\label{contour_Le100}
\end{figure}

It is interesting to note that this dominant set of modes
is not the most rapidly growing one according to linear theory:
as seen in Figure \ref{compare_gwave}, a variety of other gravity-wave modes
have equal or larger predicted growth rates, as for example ${\bf k} = (l_2, 0, k_3)$.
However, not all modes grow as predicted.
A systematic comparison between the observed and predicted growth
rates of all nine sets of modes shown in Figure \ref{compare_gwave}
reveals that mean-field theory becomes
progressively less accurate with increasing wavenumber. This information is summarised in Figure \ref{contour_Le100}. Since for our selected set of parameters a finger is typically 1 FGW wide and 5 FGW high (see Paper I),
and 1FGW $\sim 13d$, our results suggest that modes
grow as predicted only if their horizontal extent is wider than about 10 fingers
and their vertical extent is larger than about 5 fingers. Modes with intermediate
horizontal or vertical extent still grow but
slower than expected. Modes which only contain 3 or fewer fingers (horizontally or vertically) do not grow at all. These results are not surprising given that separation between the mode scale and the individual finger scale is required for mean-field theory to be applicable. Note that, when viewed in terms of finger sizes, the growth rate plot for the heat/salt case at $R_0=1.5$, shown in figure \ref{contour_Le100}b), is very similar to the one obtained for $\tau=1/3$. It therefore seems reasonable to expect that finger driven oceanic gravity waves will also occur on length scales larger than the predicted fastest growing mean-field modes.

\begin{figure}
\noindent\includegraphics[width=\linewidth]{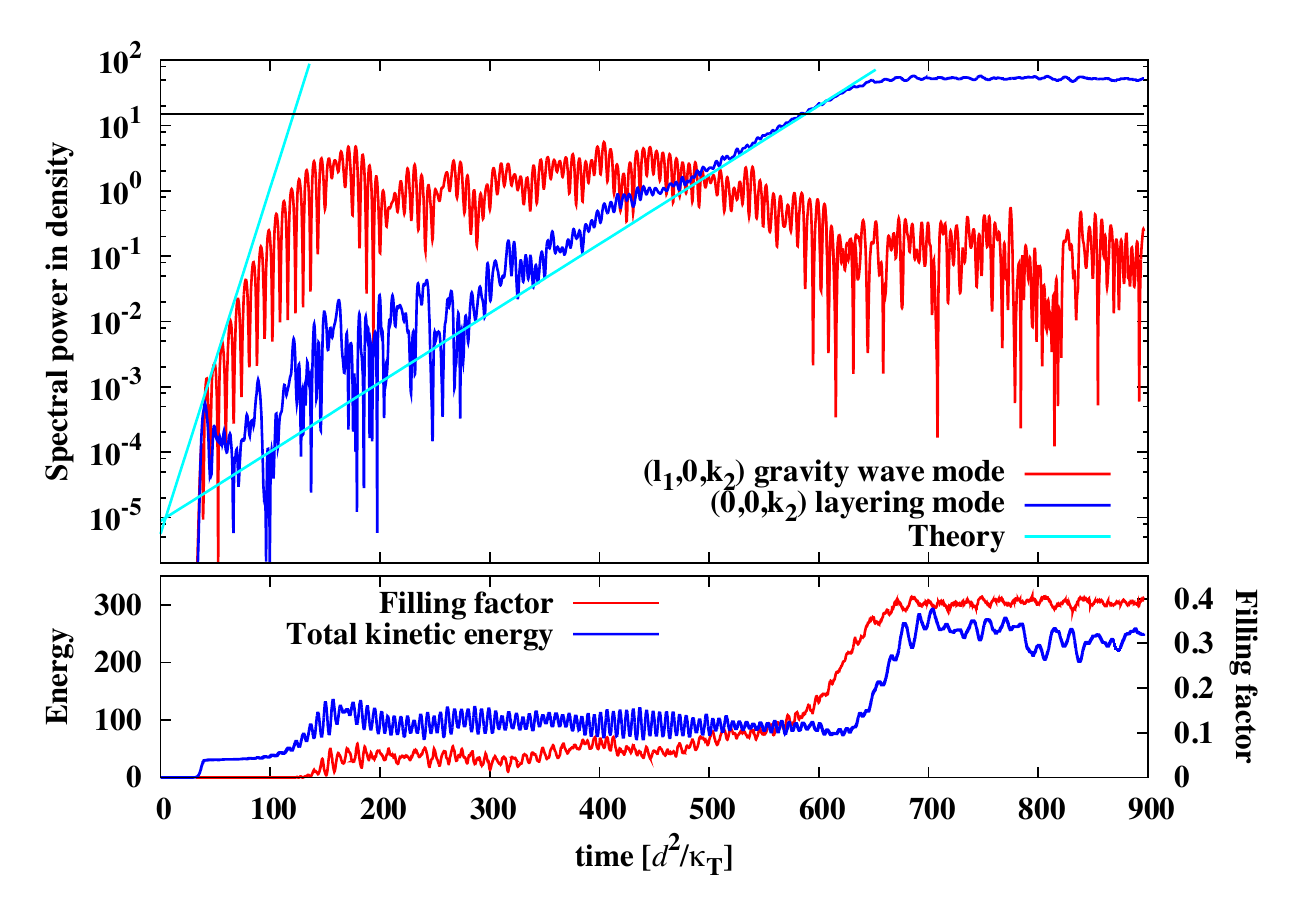}
\caption{Top: Temporal evolution of one of the dominant gravity-wave modes
($(l_1,0,k_2)$, with horizontal wavelength 335$d$, vertical
wavelength 236$d$) and of the $(0,0,k_2)$
$\gamma$-mode (vertical wavelength: 236$d$)
 respectively, as simulated numerically and
as predicted by mean-field theory (Paper I).
Shown is the norm of the Fourier coefficient of the density
perturbation (the density spectral power).
While the gravity-wave mode rapidly saturates
the $\gamma$-mode slowly but steadily grows up to the critical amplitude for
overturning (black line). Bottom: Filling factor of convectively
unstable regions (${\rm d}\rho/{\rm d}z > 0$), and total kinetic
energy of the system. Note that in order to eliminate finger-scale
density inversions, the filling factor is obtained from a low-pass filtered density field.
}
\label{fig_test_gamma_instab}
\end{figure}

Finally, Figure \ref{compare_gwave} also reveals that all
gravity-wave modes stop growing
roughly at the same time, around $t \sim 150-200$. We now show that
this is due to localised overturning events which are regularly
triggered when some of the
dominant modes constructively interfere to produce very high amplitude
perturbations in the density field.

A single gravity-wave mode with vertical wavenumber $k$
causes quasi-periodic oscillations of the total
density field (background + perturbations).
As the mode amplitude $\hat \rho_{l,m,k}$
grows, localised regions are progressively more
weakly stratified. They become unstable to direct
overturning convection when the perturbation
reaches the critical value $\hat \rho_{l,m,k} = (1-1/R_0) /k$.
Figure \ref{fig_test_gamma_instab}a compares the
spectral power in the density perturbation of one of the four dominants modes --
defined as the norm of the relevant coefficient of the Fourier
transform of the density field --
with the square of the critical amplitude for overturning. For this single
mode, the criterion for overturn is clearly never reached.  However, four of
these dominant modes co-exist and can interact constructively
to cause a localised overturning event. This effect is demonstrated in Figure
\ref{fig_test_gamma_instab}b, which shows
the filling factor of large-scale regions with positive
density gradient, or in other words the relative volume of regions
unstable to overturning convection. As expected, this filling factor is close to
zero in the homogeneous fingering phase. It
then rapidly increases to a few percent
around $t=150$, which corresponds to the time when the
density spectral power of one of the dominant gravity-wave modes
reaches about a tenth of the critical amplitude
for overturning.
After $t=150$, the steady conversion of kinetic energy
into potential energy by the localised and intermittent overturning events
causes {\it all } the gravity-wave modes to stop growing and not just
the dominant ones (see Figure \ref{compare_gwave}).
This result is not surprising given the nonlinear nature of the
breaking events.

Our results show that the gravity wave-field saturates at fairly high amplitude,
but {\it without directly triggering the formation of thermohaline layers}. The
gravity-wave-dominated phase continues on for many more oscillation
periods (from $t\sim 200$ to $t\sim 600$) and the movie of the
simulations (see online material) shows no obvious visible change in the statistical
properties of the flows. Our simulations show that the collective instability, as was originally proposed,
does not appear to be responsible for the transition to the layered state
observed in Phase III.

\subsection{Phase III : Layered phase}

\begin{figure}
\centerline{\noindent\includegraphics[width=\linewidth]{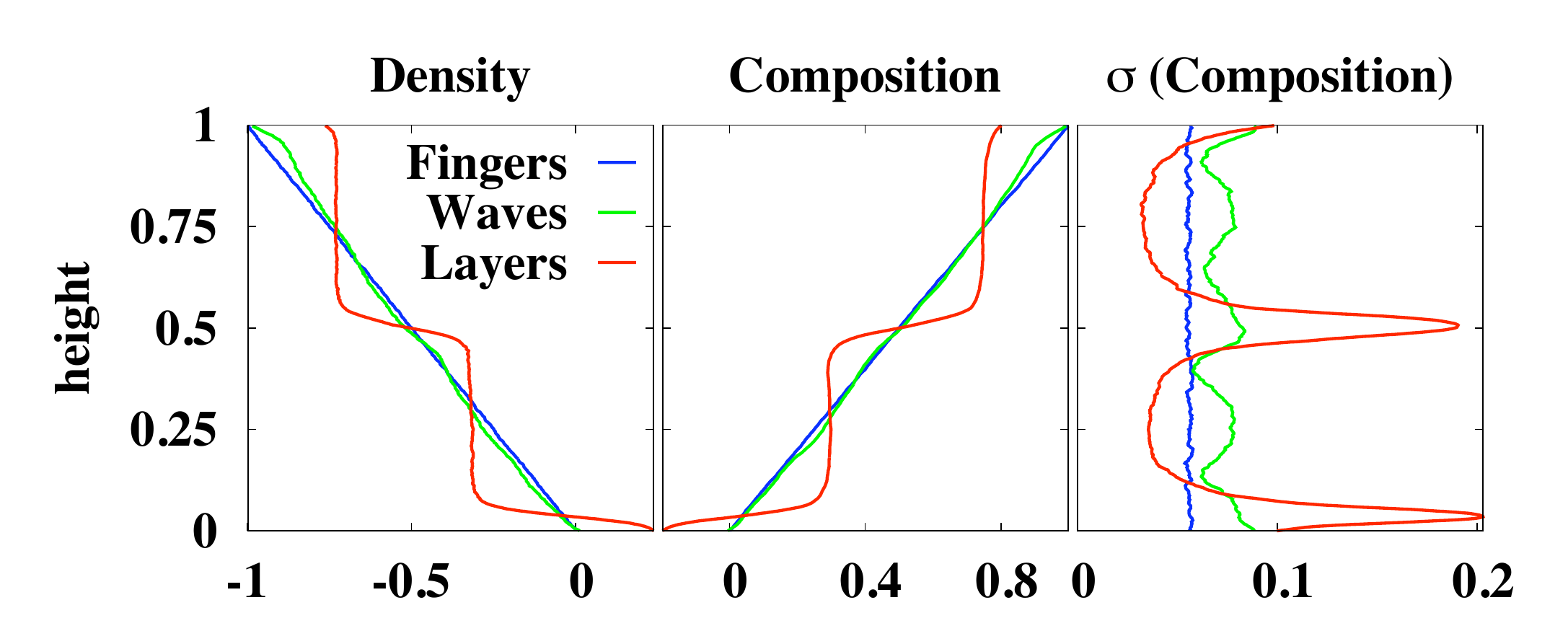}}
\caption{Horizontally averaged total density and salinity (in both cases, background + perturbation)
  and standard deviation of salinity $\sigma(S)$.
  Density and salinity are normalised by the respective
  background contrast over the entire domain, and height is given in
  units of the domain height. The three lines correspond to
  the visualisations in Figure \ref{figure_simulated_layering}. In the fingering phase, the salinity and density
  profiles are indistinguishable from the background profiles, and the standard deviation of salinity is
  statistically homogeneous. In the gravity-wave phase, the profiles remain close to the background profiles,
  but the presence of a $(0,0,k_4)$ $\gamma-$mode can be seen in the
  $\sigma(S)$ profile.
The layered phase exhibits a clear staircase in the salinity and
density profiles, while $\sigma(S)$ shows
 strong peaks corresponding to the position of the fingering interfaces.}
  \label{figure_vertical_profiles}
\end{figure}

The gravity-waves dominated phase suddenly
ends around $t \sim 600$, with the spontaneous emergence of
two convecting layers separated by thin
fingering interfaces (Figure \ref{figure_simulated_layering}c).
This layering transition is accompanied by a significant increase in the
global heat and salinity fluxes (Figure
\ref{figure_simulated_layering}d),
as well as in the total kinetic energy (Figure \ref{fig_test_gamma_instab}).
Figure \ref{figure_vertical_profiles} shows the resulting staircase in
the mean profiles: turbulent mixing leads to a nearly homogeneous
temperature and composition in the bulk of the layers, while the
strongly-stratified  interfaces exhibit larger spatio-temporal
fluctuations in the temperature and
salinity (as measured by the standard deviation from the mean) caused
by the remaining fingering. The staircase, once formed, is extremely
robust and persists despite being occasionally pierced by   strong
convective plumes.

\subsection{The $\gamma$-instability and the formation of thermohaline staircases}

In order to understand what causes the layering transition, we
now study Phase II again
in the light of predictions from mean-field theory. This time, we focus on the
$\gamma$-modes discussed in Paper I and in Section \ref{sec:intro}.
By virtue of being horizontally invariant and vertically periodic,
$\gamma$-modes might straightforwardly overturn into a fully-formed
thermohaline staircase once they reach the same critical amplitude as the one
discussed in the context of gravity-wave modes above (see Section \ref{sec:phase2}).

Since the emergent staircase in our 3D simulation has two layers,
we naturally begin by studying the $(0,0,k_2)$ $\gamma$-mode.
Figure \ref{fig_test_gamma_instab} shows that the $(0,0,k_2)$
mode is indeed present
in the simulation, and began to grow at the same time ($t \sim 50-100$) as the gravity waves
studied above. Remarkably, we find that its growth rate is well modelled by linear
mean-field theory all the way to about $t=600$, even though the mean field approach
does not account for the turbulent mixing induced by the overturning waves.
This result establishes the predictive power of our formalism.

Figure \ref{fig_test_gamma_instab} shows that this $\gamma-$mode reaches
the critical amplitude for the onset of convective overturning at $t \sim 580$
and stops growing shortly thereafter when a regular staircase in the density
(or temperature and salinity) profile appears ($t \sim 600$).
The layers rapidly become fully convective, as can be
seen in the strong increase in the filling
factor of convectively unstable regions and of the total kinetic energy
(see Figure \ref{fig_test_gamma_instab}), and the corresponding
increase of the Nusselt number
(see Figure \ref{figure_simulated_layering}d). Finally, note that the gravity waves
can no longer exist in the layered state. Their amplitudes are strongly
reduced, and stop showing a clear oscillatory signal.

\subsection{The interaction of $\gamma-$modes and gravity-wave modes}

The formation of  thermohaline staircases in our simulation thus
appears to depend only on the growth and eventual convective
overturning of $\gamma-$modes, in a way which can be understood
quantitatively using the mean-field theory and small-scale flux laws
presented in Paper I and in Figure 1. In many ways, the fact that mean-field theory holds
for the staircase progenitor, the $(0,0,k_2)$ $\gamma-$mode, is rather remarkable given the presence of the strong gravity wave field, and deserves further attention. Moreover, as shown by Radko (2003),  the $\gamma$-instability
suffers from an ultra-violet catastrophe whereby the mode growth rates
increase quadratically with wavenumber. As a result, the $(0,0,k_2)$ $\gamma-$mode is not the most rapidly growing one according to linear theory, and one may wonder what led to its selection as the staircase progenitor.

\begin{figure}
\centerline{\includegraphics[width=0.95\linewidth]{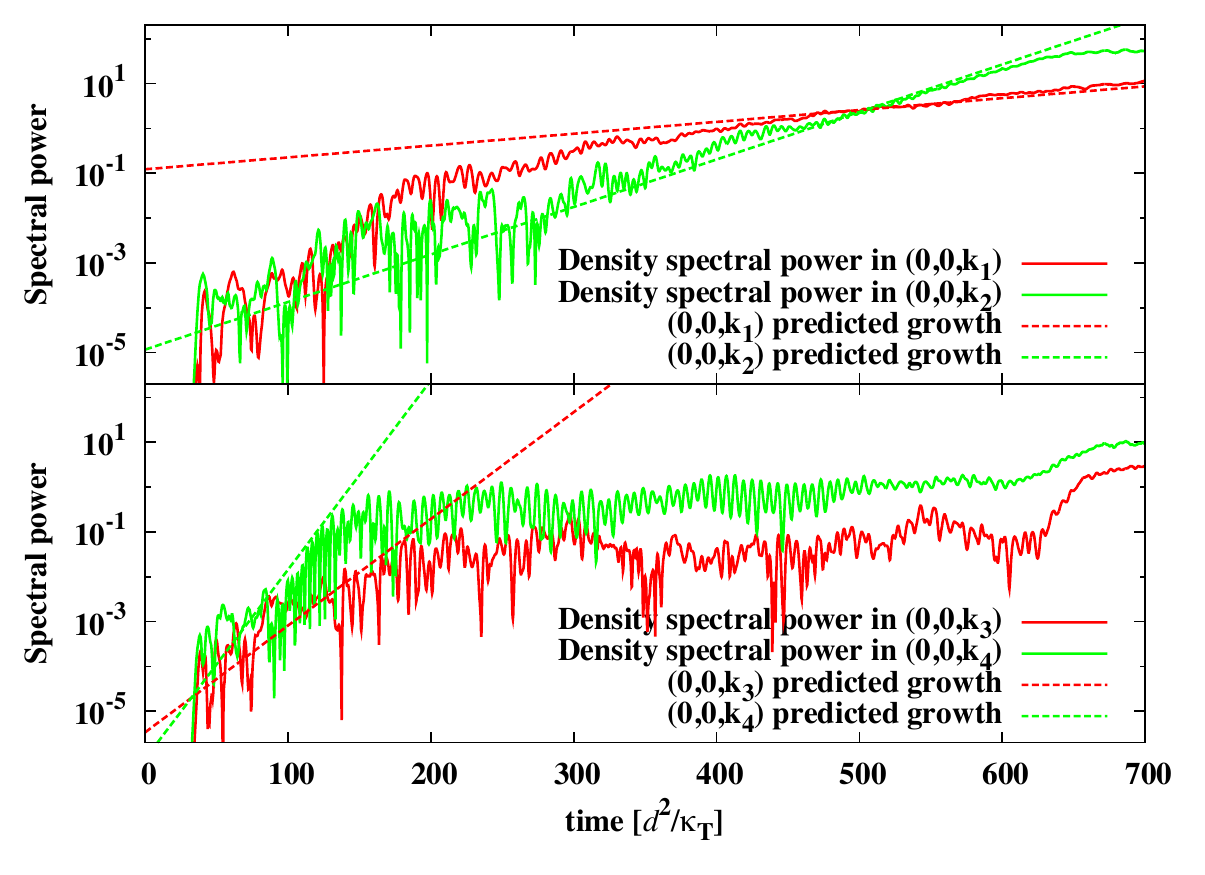}}
\caption{Density spectral power in the four largest vertical modes
  (with respective vertical wavenumbers $k_1, k_2, k_3$ and $k_4$),
  compared with mean-field theory. The two gravest modes (top panel)
  continue growing throughout the gravity-wave-dominated phase, while
  the other two modes (lower panel) stop growing around $t \simeq
  150-200$.}
\label{compare_ks}
\end{figure}

To answer these questions, we begin by extracting information on other $\gamma-$modes in the
system, with vertical wavenumbers ranging from $k_1$ to $k_4$ (modes with
higher wavenumbers, as discussed earlier, are poorly represented by mean-field theory).
Their temporal evolution is shown in Figure \ref{compare_ks} and
compared with theoretical expectations.
As expected, the $(0,0,k_4)$ mode grows the fastest among the ones studied,
followed by $(0,0,k_3)$, $(0,0,k_2)$ and finally $(0,0,k_1)$.
At early times ($t<200$), we see that the measured growth rates are
indeed close to the
corresponding theoretical ones for all modes except for $(0,0,k_1)$,
which unexpectedly grows significantly faster than predicted --
the origin of this discrepancy remains a puzzle.

\begin{figure}
\centerline{\includegraphics[width=\linewidth]{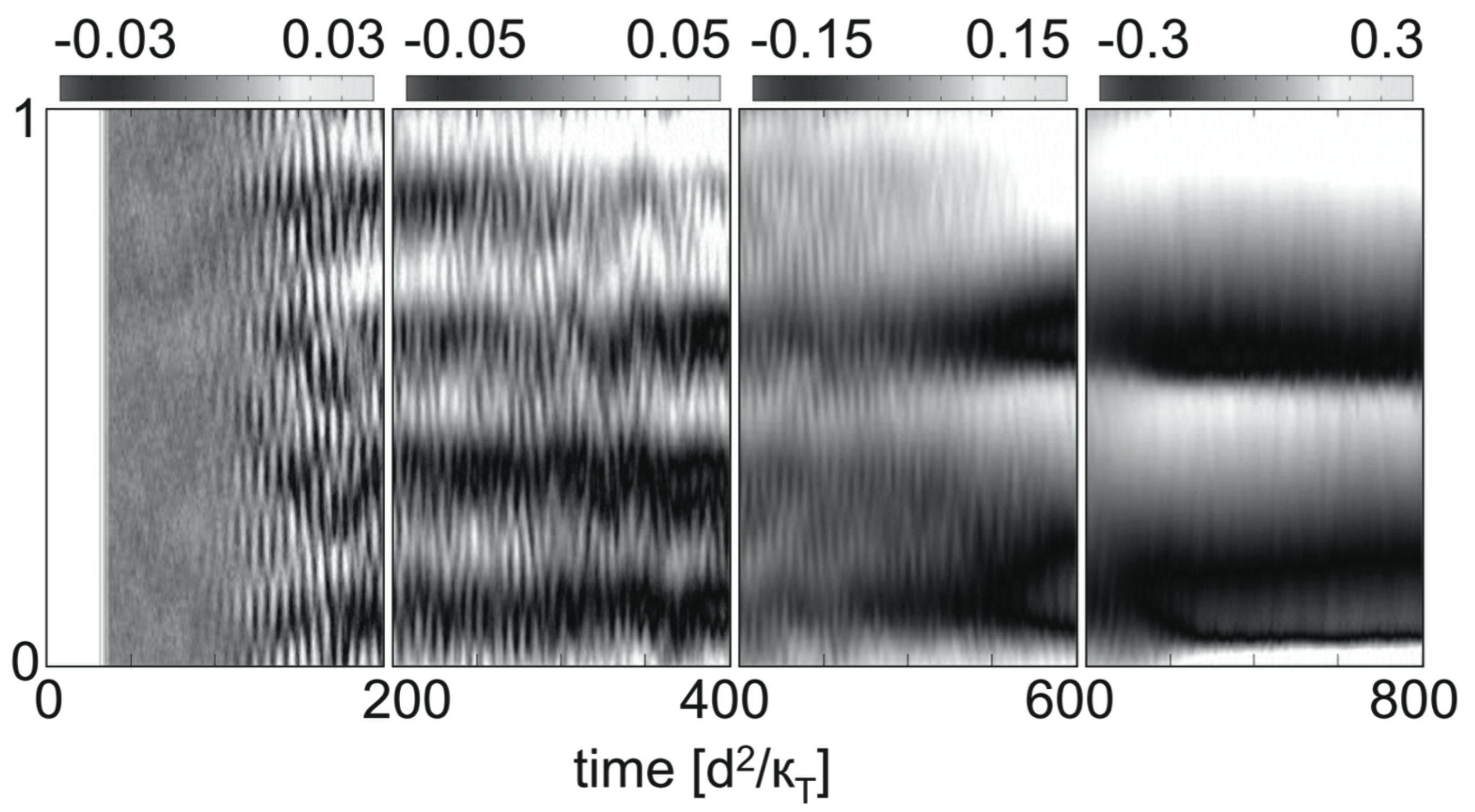}}
\caption{Horizontally averaged density perturbation
  $\langle\rho\rangle_h(z,t)$ as
a function of time from $t=0$ to $t=800$. Four intervals are shown,
each on a different colour scale to provide adequate contrast. From $t=0$ to
$t=100$, mean perturbations to the background density are close to
zero. However, notably regular oscillations appear around $t=100$, with a $k_4$
vertical wavenumber and twice the temporal frequency of the dominant gravity-wave mode.
By $t=200$, the system shows clear evidence for the presence of a
$(0,0,k_4)$ $\gamma-$mode, strongly modulated by the gravity-wave
field. Around $t=500$, the $(0,0,k_2)$ mode
emerges as the dominant mode, continues to grow and begins to overturn
around $t=600$ into two mixed layers separated by sharp interfaces. }
\label{meandensity}
\end{figure}

The subsequent evolution of the modes, beyond $t=200$ in Figure \ref{compare_ks},
reveals an interesting dichotomy
between the two larger-scale modes (top panel) which both grow
with the expected growth rate, and the two smaller-scale modes (lower panel)
which stop growing at about $t=150-200$, when the gravity waves saturate.
The amplitude of
the $(0,0,k_4)$ mode remains the largest one until about
$t=500$, even though it stopped growing at early times.
Had it continued growing only slightly further, this mode would have
been the one to cause the staircase formation, resulting in an initial layer
height half of the one observed in the simulation.
The evolution of the horizontally averaged
density perturbation (denoted as $\langle\rho\rangle_h(z,t)$ from here on)
shown in Figure \ref{meandensity} confirms more visually the dominance
of the  $(0,0,k_4)$ mode until about $t=500$,
and the transition to a $(0,0,k_2)$ mode thereafter (with a hint of $(0,0,k_1)$
inducing a small asymmetry between the two layers).

The results of Figure \ref{compare_ks} naturally prompt us
to wonder why the smaller-scale
modes stop growing during the gravity-wave phase
while the larger-scale, slower growing modes do not.
This important question impinges on the problem of
the selection of the $\gamma-$mode which eventually becomes the staircase progenitor,
and thus determines the initial layer spacing.
In order to proceed further and understand what may cause the smaller-scale
modes to stop growing, we need to gain further insight on
the nonlinear dynamics of the various mean-field modes.

We can identify the dominant nonlinear mechanism for interaction
between the various mean-field modes with a closer inspection of the
$\langle \rho \rangle_h(z,t)$ profile.
Figure \ref{meandensity} reveals notably regular
oscillations from $t=100$ to $t=600$ which
can be traced back to the nonlinear buoyancy transport induced by the
dominant gravity-wave modes studied in Section \ref{sec:phase2}. Indeed, a
gravity-wave with vertical wavenumber $k$ and frequency $\omega_g$ induces a horizontally averaged
vertical heat (or salt) flux $\langle wT \rangle $ or $\langle wS \rangle$ proportional to $\exp(2ikz + 2i\omega_g
t)$. Since the dominant gravity-wave mode has a $k_2$ vertical wavenumber, the induced
perturbations in $\langle \rho \rangle_h$ are expected to have a $k_4$
wavenumber and to oscillate at twice the wave frequency. This oscillation is indeed
observed, and is most regular {\it before} the wave-breaking events
begin, between $t=100$ and $t=150$ in Figure \ref{meandensity}.

By virtue of being a quadratic quantity in the wave amplitude, the wave-induced flux is neglected in
any standard linear mean-field theory. However, Figure \ref{meandensity} shows that it remains
important throughout the entire wave-dominated phase from $t=100$ to $t=600$.
In particular, it is responsible for the oscillations
which are seen to modulate the $(0,0,k_4)$ $\gamma-$mode rather
strongly, either enhancing it or suppressing it depending on the phase of the wave.
One may therefore conjecture that the $(0,0,k_4)$ (and presumably the $(0,0,k_3)$) $\gamma-$modes
stop growing as a result of nonlinear interactions with the dominant
gravity-wave modes through the wave-induced flux.
Unfortunately, a complete quantitative understanding of the
numerical results from our simulations, in particular in view of
answering the question raised above, can only be gained from a
thorough nonlinear analysis of our mean-field equations, a problem
which is beyond the scope of the current paper. However, given that the
dominant gravity-wave mode induces horizontally averaged
fluxes with exactly the same vertical structure as that of the $(0,0,k_4)$ $\gamma-$mode,
the conjecture seems reasonable.

The question then remains of why the $(0,0,k_2)$ and $(0,0,k_1)$ $\gamma-$modes on the other hand still
grow at the predicted rate despite the effect of the wave-induced flux. This result is even more
remarkable for the $(0,0,k_2)$ mode, which has an amplitude several orders of magnitude smaller
than that of the dominant gravity-wave modes at early times. Again, without a
full analysis of the nonlinear mean-field equations we are left to speculate on the matter.
A plausible answer lies in the spatial mismatch between the small-scale
flux modulation induced by the waves (working on the $k_4$ lengthscale) and the intrinsically larger scale of the two gravest $\gamma-$modes. However, showing this
would again require a weakly nonlinear analysis of the mean-field equations, and is deferred to future work.

The results of the simulations, however, stand beyond the
speculations. Whatever the cause for this observed effect, it appears
that the dominant gravity waves act as a ``filter'' for any
$\gamma$-mode with significantly smaller vertical wavelength. If we
generalize this result to other parameter regimes, the progenitor of
the staircase is always likely to be a $\gamma-$mode with wavenumber
commensurate with that of the dominant
gravity waves excited by the collective instability. As such, it is
interesting to note that both sets of modes play a role in the
eventual formation of a thermohaline staircase,  although
in a way which was neither anticipated by \citet{stern1969cis}, nor by \citet{radko2003mlf}.

\section{Discussion and prospects}
\label{sec:ccl}

Our simulation, when combined with the results of
mean-field theory derived in Paper I, presents a clear and simple
view of the sequence of events which may lead to the spontaneous formation of  oceanic
staircases (in the absence of lateral gradients or other sources of mechanical mixing), and
which can also be straightforwardly generalised to other double-diffusive systems.

Let us first consider a homogeneous heat-salt system, with constant
background temperature
and salinity gradients unstable to fingering convection. At low enough density ratio ($R_0 < 4$, see Paper I),
the turbulent flux ratio $\gamma$ induced by fingering convection
decreases most rapidly with $R_0$.
This regime is simultaneously unstable to two types of secondary large-scale ``mean-field'' instabilities: the collective
instability which excites gravity waves \citep{stern1969cis,stern2001sfu},
and the $\gamma-$instability which leads to the
growth of horizontally-invariant perturbations in the temperature and salinity field \citep{radko2003mlf}.

As shown in Paper I, mean-field theory predicts that gravity-wave modes grow more rapidly than $\gamma$-modes for values
of $R_0$ characteristic of regions of the ocean where thermohaline staircases are observed (i.e. $1 < R_0 < 1.8$).
By comparison with numerical simulations at $\pran =7$ and $\tau = 1/3$ we also found that the theoretically fastest-growing gravity-wave modes are actually too small for mean-field theory to be applicable. Instead,  the emerging dominant gravity-wave modes are slightly more extended and have smaller
theoretical growth rates. If we extrapolate the results of Figure
\ref{contour_Le100}a to parameter regimes relevant of
the heat-salt system (see \ref{contour_Le100}b), taking
$R_0 = 1.5$ as a typical value for staircase regions, we find that the
overall picture is very similar:
modes which are large enough to grow with the theoretically predicted
rate are very slowly growing modes, while the theoretically fastest
growing ones are too small for mean-field theory to be applicable. As
a result, we expect the dominant gravity-wave modes in the heat-salt system to be
intermediate in size (20-30 fingers in width and 4-5 fingers in
height), and grow with a non-dimensional growth rate of the order of
about 0.04-0.05, taking into account that their real growth rate is
expected to be smaller than the theoretically predicted one. Putting these results in
dimensional form, using typical oceanic values of $\nu=10^{-6}$m$^2$s$^{-1}$,
$\kappa_T= 1.4\times 10^{-7}$m$^2$s$^{-1}$, $\alpha_T= 2\times 10^{-4}
$K$^{-1}$, $g=10$ms$^{-2}$ and $T_{0z}=10^{-2} $Km$^{-1}$, we
have $d \sim 0.9$ cm and 1FGW = $8.2d$, so that a typical finger
is of the order of 7 cm wide and 35 cm high. Hence the gravity
waves expected to dominate the dynamics
have horizontal and vertical  wavelengths of the order
of 1.5m - 2m, and grow on a timescale of the order of a few (3-5) hours.

These gravity waves saturate rapidly as a
result of localised breaking events, without directly inducing layer
formation. Meanwhile, a spectrum of $\gamma$-modes also grow, with one of them
destined to become the progenitor of the thermohaline staircase.
Our results suggest that the dominant gravity-wave modes interact nonlinearly
with the $\gamma$-modes and act as a filter for the smaller-scale ones. Only those
with vertical wavelengths larger than that of the dominant gravity-wave modes
are allowed to grow, but then appear to do so on a timescale well-predicted
by mean-field theory. Since the fastest growing $\gamma$-modes are those with the smallest vertical
wavelength, we expect the progenitor of the staircase to be a mode
with vertical spacing commensurate with the vertical wavelength of the
dominant gravity-wave modes, i.e. around 5 fingers in height or about
1.5m-2m for parameters appropriate of the heat-salt system.
This mode grows on a timescale of 5-6 hours, and upon reaching a critical
amplitude,
causes regularly-spaced inversions in the mean density profile which are subject
to direct overturning convection. A regular staircase then forms, with an
initial layer spacing of about 2m.
Note that once formed, experiments \citep{krishnamurti2003ddt}
and theory \citep{radko2003mlf,radko2005dtl,radko2007mme} suggest
that the staircase evolves further in time through successive mergers
until an equilibrium layer depth is established. These events have not
been observed in our 3D simulations yet since integrating the simulation on a merger timescale \citep{radko2005dtl} is numerically prohibitive in 3D.

Beyond giving new insight into the origin of oceanic staircases, our
study marks a starting point in answering the question of whether
gravity modes and $\gamma$-modes can be excited in other less
readily observable fluid systems as well, and whether staircases may
be expected or not. For example, it has been pointed out that vigorous
fingering convection might occur in stars
\citep{vauclair2004mfa,charbonnel2007tmp,stancliffe2007cem},
where the resulting transport may account for observed peculiarities
in luminosity or metallicity. Our work suggests that, despite the
complexity of these highly nonlinear processes,  much can still be
learned first by obtaining local flux laws for turbulent mixing in
the parameter regimes typical of these systems, and then applying the
mean-field theory framework as show in Paper I and in this work.

\begin{acknowledgments}
A.T., P.G.  and T.R. are supported by the National Science Foundation,
NSF-093379, and T.R. is supported by an NSF CAREER. S.S. was supported by grants from the NASA Solar and Heliospheric Program (NNG05GG69G,NNG06GD44G,NNX07A2749). The simulations
were partially run on the Pleiades supercomputer at UCSC, purchased using an
NSF-MRI grant. Computing time was also provided by the John von Neuman Institute for Computing. We thank Gary Glatzmaier for many helpful discussions and for his continuous support.
\end{acknowledgments}

\bibliographystyle{jfm}
\bibliography{DDC_bib}

\end{document}